\documentstyle[floats,aps,prl,twocolumn,epsf]{revtex}

\newcommand{\be}{\begin{equation}}
\newcommand{\ee}{\end{equation}}
\newcommand{\bea}{\begin{eqnarray}}
\newcommand{\eea}{\end{eqnarray}} 
\newcommand{\vap}{\varepsilon}

\tightenlines 
 
\begin{document} 

\bibliographystyle{unsrt}

\title{The Universal Negative Poisson Ratio of Self-Avoiding
Fixed-Connectivity Membranes}

\author{M. Bowick,$^{1}$ A. Cacciuto,$^{1}$ G. Thorleifsson$^{2}$ and 
A. Travesset$^{3}$} 
 
\address{$^1$ Physics Department, Syracuse University,  
Syracuse NY 13244-1130, USA \\} 
 
\address{$^2$ DeCODE Genetics, Lynghalsi 1,IS-110, Reykjavik, Iceland\\} 
 
\address{$^3$ Loomis Laboratory, University of Illinois at Urbana, Urbana IL 
61801, USA\\}   
 
\maketitle 
 
\begin{abstract} 
 
We determine the Poisson ratio of {\em self-avoiding} fixed-connectivity 
membranes, modeled as impenetrable plaquettes, to be  $\sigma=-0.37(6)$, 
in statistical agreement with the Poisson ratio of {\em phantom} fixed-connectivity 
membranes $\sigma=-0.32(4)$. Together with the equality 
of critical exponents, this result implies a {\em unique} universality  
class for fixed-connectivity membranes. 
Our findings thus establish that physical fixed-connectivity membranes provide a 
wide class of {\em auxetic} (negative Poisson ratio) materials with significant  
potential applications in materials science. 
 
\end{abstract} 
 
\normalsize

Fixed-connectivity (also known as polymerized, tethered or crystalline) 
membranes are fluctuating and flexible fishnet-like 
two-dimensional surfaces with nodes of fixed coordination number 
(for two recent reviews see \cite{BowTra:00,Wiese:00}). 
Physical examples include such naturally occurring  
structures as polymerized Langmuir-Blodget films 
\cite{Fendler1,Fendler2} and the spectrin/actin cytoskeleton 
\cite{Skel:93} of erythrocytes (mammalian red blood cells). 
A wide variety of additional examples is discussed in 
\cite{BowTra:00}. 
Current advances in soft condensed matter  
experimental techniques suggest the likelihood of many new 
realizations of fixed-connectivity membranes such as cross-linked DNA 
networks as well as composite structures that include fixed-connectivity 
membranes as fundamental ingredients. One universal and remarkable feature 
of the low-temperature (so-called {\em flat}) phase of 
non-self-avoiding ({\em phantom}) fixed-connectivity membranes is that  
they expand transversely when stretched 
longitudinally \cite{AL:88,LDR:92,ZDK:96,FBGT:97}. In other words,
they exhibit a negative Poisson ratio \cite{Landau7}.
Such materials have been dubbed {\em auxetic} \cite{EVNK:91}.

In this letter we estimate, via Monte Carlo simulations, the Poisson 
ratio of physical {\em self-avoiding} fixed-connectivity membranes.  
We establish that they are {\em also} auxetic materials with a Poisson ratio 
and roughness exponent in statistical agreement with those of flat phantom 
membranes. Thus there appears to be a unique universality class
of flat fixed-connectivity membranes, whether they arise from high
bending rigidity or self-avoidance. 
Direct experimental measurements should be able to measure this negative 
Poisson ratio and test universality.
The remarkable properties of auxetic membranes suggest a 
rich new avenue of exploration in materials synthesis. 

A fixed-connectivity membrane may be modeled as an 
elastic surface with bending rigidity and self-avoidance 
\cite{KKN:86,NP:87,KKN:87,PKN:88}. The free energy is composed of an
elastic and a self-avoiding contribution: 

\begin{eqnarray} 
\label{full_energy}
F_{fcm} = F_{\em el} + F_{\em sa} \ . 
\end{eqnarray} 
The elastic free-energy $F_{\em el}$ is given by
\begin{eqnarray} 
\label{elastic_energy}
F_{\em el}({\bf u},h) = \int d^2{\bf x}\left[\frac{\kappa}{2} 
(\partial_{\alpha}\partial_{\beta} h)^2+\mu u_{\alpha \beta}^2  
+ \frac{\lambda}{2}u_{\alpha \alpha}^2\right]  
\end{eqnarray}
where ${\bf u}$ denotes phonon modes, $h$ denotes height modes,
$\kappa$ is the bending rigidity and $\lambda$ and $\mu$ are 
the classical Lam\'e coefficients.  
The self-avoiding free-energy we take to be of the Edwards form  

\begin{eqnarray}
\label{sa_energy}
F_{\em sa} = \frac{b}{2} 
\int d^2{\bf x} \int d^2{\bf y} \delta^{3}(\vec{r}({\bf x})-\vec{r}({\bf y})) \ , 
\end{eqnarray} 
where $b$ determines the strength of self-avoidance. 
The strain tensor $u_{\alpha \beta}$ \cite{Landau7}
is related to the embedding ${\vec r}({\bf x})$, defining the membrane by 
\bea\label{strain_tensor}  
{\vec r}({\bf x})&=&{\bf x}+{\bf u}({\bf x})+\hat{z} h({\bf x}) 
\\\nonumber 
u_{\alpha \beta}&=&\frac{1}{2}(\partial_{\alpha} u_{\beta}+ 
\partial_{\beta} u_{\alpha}+\partial_{\alpha} h \partial_{\beta} h) 
\eea 
 
\begin{figure}[ht] 
\epsfxsize= 3in \centerline{\epsfbox{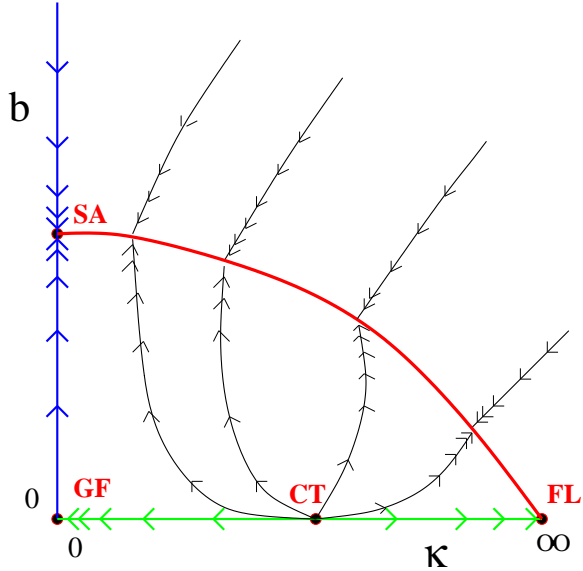}} 
\caption{The renormalization group flows in the two--dimensional space of couplings for
a fixed-connectivity membrane with bending rigidity ($\kappa$) and self-avoidance ($b$).
The phantom model ($b=0$) has two infrared stable 
fixed points, the crumpled phase (GF) and the flat phase (FL), with an  
intermediate continuous crumpling transition associated with the infrared-unstable
fixed point (CT). The pure self-avoiding model with
no microscopic bending rigidity ($\kappa=0$) has a infrared stable 
self-avoiding fixed point (SA). There is a line of 
equivalent fixed-points joining the FL and the SA (the solid (red) line), 
thus defining a redundant direction in $\kappa-b$ space.}
\label{fig__phasediagr} 
\end{figure}

Combined efforts both on the analytical and numerical side have led to 
a complete clarification of the phase diagram of the {\em phantom} case 
($b=0$) and detailed estimates for the critical exponents, as shown  
in Fig.~\ref{fig__phasediagr} and Table~\ref{tab__all}. The phase diagram consists 
of a crumpled phase (associated with the Gaussian fixed point GF)
and a flat phase (associated with a flat phase fixed-point FL), 
with an intermediate infrared-unstable crumpling transition (CT),
as depicted in Fig.~\ref{fig__phasediagr}. 
 
The self-avoiding model of Eq.~(\ref{full_energy}) with no bending 
rigidity ($\kappa=0$) has proven to be tractable  
numerically \cite{BCTT:2000}. The model possesses a unique infrared fixed point  
(SA) describing a flat phase, with detailed results 
for critical exponents given in Table~\ref{tab__all}. From the 
analytical standpoint there is evidence for a unique SA 
\cite{Dup:87,KN3:87}, but the calculations are inconclusive on whether 
it describes a flat or a crumpled phase 
\cite{Hwa:90,GuPa:92,Gou:91,LeD:92,DaWi:96,WiDa:97}.  
 
A look at Table~\ref{tab__all} reveals that the critical exponents and the  
Poisson ratio of both phantom and self-avoiding membranes coincide 
within the error bars quoted, implying that the FL and the SA  
are equivalent. This is a very surprising result since it means that 
the same long wavelength limit is reached via two very different 
routes: either sufficiently large bending rigidity or strong 
self-avoidance in the absence of bending rigidity. This is even more 
remarkable considering the very different short-distance structure of 
the two models: a membrane with large bending rigidity is very smooth 
at short distances while a purely self-avoiding membrane in the 
absence of bending rigidity is extremely rough. 
 
\begin{table} 
\begin{tabular}[tbc]{c|l|l|l|llll} 
\multicolumn{1}{c}{}  & \multicolumn{4}{c}{\bf{PHANTOM}} & 
\multicolumn{2}{c}{} & \multicolumn{1}{c}{} \\\hline 
\multicolumn{1}{c}{}  &\multicolumn{1}{c}{MC} & 
\multicolumn{1}{c}{$\vap$-expansion} &  
\multicolumn{1}{c}{SCSA} & \multicolumn{1}{l}{Large d }  & 
\multicolumn{1}{c}{}\\\hline 
$\nu$    & $0.95(5)$\cite{BCFTA:96}    & $1$  &  $1$     & $1$   &  
$$ \\\hline 
$\zeta$  & $0.64(2)$\cite{BCFTA:96}    &  $13/25$ \cite{AL:88}  &   
$0.59$ \cite{LDR:92}   & $2/3$ \cite{DG:88} & \\\hline 
$\sigma$ & $-0.32(4)$\cite{FBGT:97}  &  $-1/5$\cite{AL:88}   &  
$-0.33$\cite{AL:88}  & $$ & $$\\\hline       
\multicolumn{3}{c}{} & \multicolumn{3}{c}{} \\\hline 
\multicolumn{3}{c}{} & \multicolumn{3}{c}{} \\\hline 
\multicolumn{3}{c}{\bf{SELF-AVOIDING}} & \multicolumn{5}{c}{\bf{EXPERIMENTS}} \\\hline 
\multicolumn{2}{c}{MC-BS}&\multicolumn{1}{c}{MC-IP} \\\hline 
$\nu$&1\cite{AN2:90}&0.97(4) \cite{BCTT:2000}& \multicolumn{1}{c}{} & 0.93(5)\cite{Zasa:94} \\\hline 
$\zeta$& $0.65 \cite{AN2:90}$ &0.64(2) \cite{BCTT:2000}& \multicolumn{1}{c}{} &  0.65(10) \cite{Skel:93} && \\\hline 
$\sigma$& &-0.37(6) &\multicolumn{1}{c}{} & 
\end{tabular} 
\vspace{.25cm} 
\caption{Critical exponents and Poisson ratio of flat fixed-connectivity 
membranes in the phantom (MC:Monte Carlo; SCSA:Self-consistent screening approximation) 
and the self-avoiding case (MC-BS:Monte Carlo with Balls and Springs models; 
MC-IP:Monte Carlo with impenetrable models).}
\label{tab__all} 
\end{table} 
In light of our results, we suggest the renormalization group (RG) flows depicted in 
Fig.~\ref{fig__phasediagr}. Given these flows and the equivalence 
of the SA and the FL, there must be a full line of equivalent fixed points 
$b(\kappa)$ joining them, with no RG flow on the line (corresponding to a marginal direction). 
Since these   
are the infrared stable fixed points of the system, all the relevant physics is  
described by this line. The crumpling transition present 
for phantom membranes may be reached only by an extremely precise tuning of the  
parameters involved in the problem. This would make the crumpling transition in the model
described very difficult to verify experimentally.

We see that the combination of fixed-connectivity (integrity of the 
lattice) and self-avoidance sufficiently restricts the entropy of 
crumpled configurations as to destroy the crumpled phase. It was 
already observed in \cite{AN1:90} that next-to-nearest neighbor 
self-avoidance, discretized by hard-sphere potentials, induces a 
positive bending rigidity. On the other hand the impenetrable 
plaquette model treated here is very flexible, since only strictly 
self-intersecting configurations are forbidden, and hence the physical 
mechanism flattening the membrane is clearly more general than the simple 
induced bending rigidity discussed above.   
 
We performed a Monte Carlo simulation of a suitable discrete version 
of the model (Eq.~(\ref{full_energy})), as described in \cite{BCTT:2000}. 
The Poisson ratio $\sigma$ for a two-dimensional system deformed from its mean length 
$l$ by $\delta l$ is determined by  
\be\label{poisson} 
\sigma= - \frac{\delta w/w}{\delta l/l} \ , 
\ee 
where $\delta w/w$ measures the fractional change in the transverse extent (width) 
of the system.  
In \cite{FBGT:97} it is shown that linear response theory gives   
\be\label{Poisson_ratio} 
\sigma=-\frac{\langle \overline{u_{xx}} \overline{u_{yy}}\rangle_c} 
             {\langle \overline{u_{yy}}^2 \rangle_c} \ , 
\ee 
\noindent 
where $\langle u^2 \rangle_c$ is the connected statistical average over 
Monte Carlo configurations and $\overline{u}$ is the spatial 
average over the surface. 

\begin{figure}[ht] 
\epsfxsize= 3.5in \centerline{\epsfbox{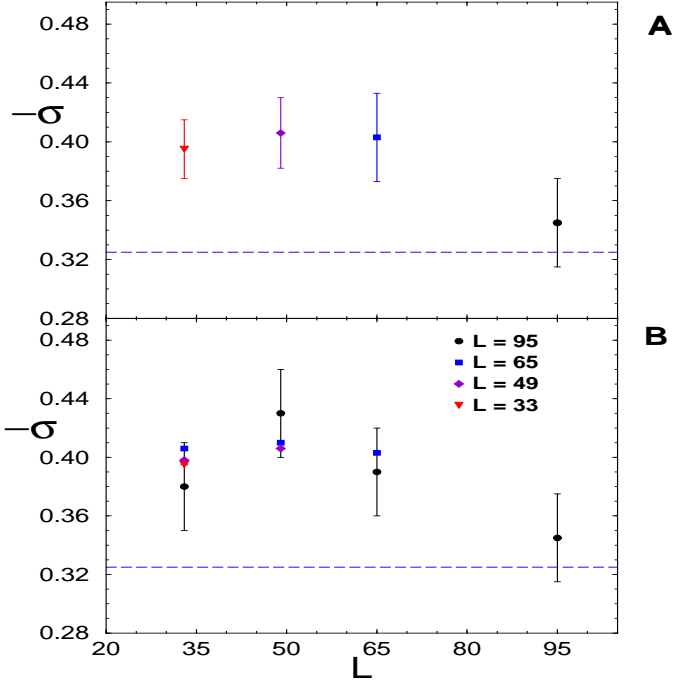}} 
\caption{Poisson ratio of a self-avoiding fixed-connectivity membrane as a 
function of system size ({\bf A}). Poisson ratio of reduced lattices compared with 
non-truncated ones ({\bf B}). The straight dashed line indicates in both cases the   
SCSA analytical result $\sigma=-1/3$.} 
\label{poiss} 
\end{figure} 
Our results are presented in Fig.~\ref{poiss}({\bf A}) and  
show that the Poisson ratio $\sigma=-1/3$ 
obtained in simulations of flat phantom fixed-connectivity membranes 
is also a good  estimate of the Poisson ratio for the pure 
self-avoiding model. 
Error bars have been calculated with the Jacknife method \cite{Efron:82,ST:95}.
 
To calibrate the influence of the boundary on our results  
we calculated the Poisson ratio 
excluding concentric outer shells of nodes neighboring the boundary. 
In particular we performed the numerical analysis discarding shells of 
boundary nodes of increasing extent. The results are presented in 
Fig.~\ref{poiss}({\bf B}). 

The exclusion of larger shells of boundary sites slightly increases the 
absolute value of the Poisson ratio. For sufficiently small reduced linear size, 
$|\sigma|$ begins to decrease due to finite size effects. 
There is consequently competition between boundary and finite size effects. 
Finite size effects become important only when a sizeable fraction of nodes 
near the boundary are excluded. 
Note that we can also make a consistent check of our analysis by
systematically excluding shells of nodes from the boundary in towards the 
center of the lattice and comparing the lattices of reduced size thus 
obtained with equal volume non-truncated lattices.  

Thus we can compare, for example, the $L=65$ result with that from  
the reduced $L=95$ lattice and likewise the $L=49$ result with that from 
the reduced $L=95$ and $L=65$ lattices. The matching given by this comparison 
is consistent. In fact we are able to reproduce the $L=65$, $L=49$ and $L=33$ 
results simply by reducing the $L=95$ lattice. 
The deviations found in this comparison are another measure of boundary effects. 

\begin{figure}[tcb] 
\epsfxsize= 3.4in \centerline{\epsfbox{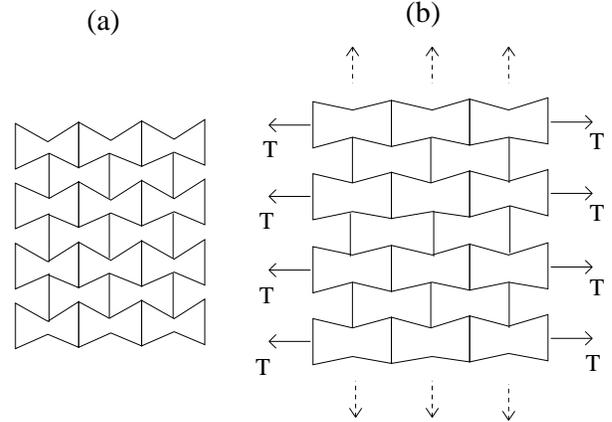}} 
\vspace{.25cm} 
\caption{Mechanical model of an auxetic material: (a) in the absence 
of applied stress and (b) under applied lateral stress $T$.  The 
lateral stretching accompanying the applied stress forces the  
material out in the transverse dimension.} 
\label{fig__aux} 
\end{figure} 
 
Traditional materials get thinner when stretched and fatter when 
squashed, since it is typically difficult to increase their volume 
very much when deformed.\footnote{A volume preserving deformation has 
Poisson ratio $0.5$. Any value less than $0.5$ involves some increase 
in volume under deformation. A negative Poisson ratio implies a very 
large volume increase.} In the unusual world of auxetics the opposite 
happens, with a number of interesting implications and potential 
applications. There are several well known auxetic materials. The 
earliest example, dating from more than a century ago, is that of a 
pyrite crystal \cite{LOV:91}, which has a Poisson ratio in certain 
directions of $\sigma \sim -0.14$. More recently some polyester foams 
under certain pressure conditions have proved to be isotropic auxetic 
materials with Poisson ratios as large as $\sigma \sim -0.7$ 
\cite{LAK:87}. A nice mechanical model for auxetic materials was 
given in \cite{EVNK:91} (see Fig.~\ref{fig__aux}).
One of the rare naturally occurring auxetics is ${\rm SiO_2}$ 
in its $\alpha$-crystobalite phase \cite{YEG:92,KES:92}.   
 
The underlying mechanism driving fixed-connectivity membranes auxetic has 
some similarities to that illustrated in Fig.~\ref{fig__aux}. 
Submitting a membrane to tension will suppress its out-of-plane 
fluctuations, forcing it entropically to expand in both in-plane 
directions. More physically, the out-of-plane undulations renormalize 
the elastic constants (the Lam\'e coefficients), in such a way that 
the long-wavelength bulk modulus is less than the shear modulus, which 
is the signature of a two-dimensional auxetic material. 
 
Auxetic materials have desirable mechanical properties such as higher 
in-plane indentation resistance, transverse shear modulus and bending 
stiffness. They have clear applications as sealants, gaskets and  
fasteners. They may also be promising materials for artificial 
arteries, since they can expand to accomodate sudden increases in blood 
flow.  

We can model a realistic fixed-connectivity membrane with an elastic free 
energy and either large bending rigidity or self-avoidance. This is of 
practical importance in modeling since, for example, we may replace 
the more complicated non-local self-avoidance term with a large 
bending rigidity. It remains an important theoretical challenge to 
verify this conclusion analytically.  
 
In this letter we have ignored the role of topological defects.  
We think that a sufficiently large defect density may affect  
the actual value of the Poisson ratio, but a detailed discussion 
of this topic is beyond the scope of this paper.  
 
It is our hope that the results presented in this letter will  
encourage materials scientists and condensed matter experimentalists
to further study the elastic and mechanical properties of 
fixed-connectivity membranes and, in particular, to measure the 
Poisson ratio of these novel systems.  
 
\bigskip 
 
The work of M.B., A.C. and A.T. has been supported by the U.S. Department  
of Energy (DOE) under contract No. DE-FG02-85ER40237. 
A.T. acknowledges funding from the materials computation center, grant
NSF-DMR 99-76550 and NSF grant DMR-0072783.

\end{document}